\documentclass[aps,prb,epsf,preprint,superscriptaddress,showpacs,preprintnumbers,12pt]{revtex4-1}
\usepackage{amsmath}
\usepackage{graphicx}

\begin{document}

\title{Impurity electrons in narrow electric field-biased armchair graphene nanoribbons}

\author{B.~S.~Monozon}
\affiliation{Physics Department, Marine Technical University, 3 Lotsmanskaya Str.,\\
190008 St.Petersburg, Russia}

\author{P. Schmelcher}
\affiliation{Zentrum f\"{u}r Optische Quantentechnologien,
Universit\"{a}t Hamburg, Luruper Chaussee 149, 22761 Hamburg,
Germany} \affiliation{The Hamburg Centre for Ultrafast Imaging,
Universit\"{a}t Hamburg, Luruper Chaussee 149, 22761 Hamburg,
Germany}

\date{\today}

\begin{abstract}
We present an analytical investigation of the quasi-Coulomb impurity states in a narrow
gapped armchair graphene nanoribbon (GNR) in the presence of a uniform external
electric field directed parallel to the ribbon axis. The effect of the ribbon confinement
is taken to be much greater than that of the impurity electric field,
which in turn considerably exceeds the external electric field. Under these conditions
we employ the adiabatic approximation assuming that the motion parallel
("slow") and perpendicular ("fast") to the ribbon axis are separated adiabatically. In the approximation of the
isolated size-quantized subbands induced by the "fast" motion the complex energies
of the impurity electron are calculated in explicit form. The real and imaginary parts
of these energies determine the binding energy and width of the quasi-discrete
state, respectively. The energy width increases with increasing the electric field
and ribbon width. The latter forms the background of the mechanism of
dimensional ionization. The S-matrix - the basic tool of study of the transport problems
can be trivially derived from the phases of the
wave functions of the continuous spectrum presented in explicit form.
In the double-subband approximation we calculate the complete widths of the
impurity states caused by the combined effect of the electric field and the Fano
resonant coupling between the impurity states of the discrete and continuous spectra
associated with the ground and first excited size-quantized subbands.
Our analytical results are shown to be in
agreement with those obtained by other theoretical approaches.
Estimates of the expected experimental values for the typically employed GNRs
show that for weak electric field the impurity quasi-discrete states remain sufficiently stable to be observed in
corresponding experiment, while relatively strong field unlock the captured electrons to further restore
their contribution to the transport.
\end{abstract}

\maketitle

\section{Introduction}\label{S:intro}

Experimental and theoretical studies of the transport, electronic and optical
properties of the armchair graphene nanoribbon (GNR) have attracted much attention
in recent years. One of the reason for this is that the GNRs
used as the interconnects in graphene-based nanoelectronic
and as the basic elements in the logic transistors could provide ultrahigh
carrier mobility between the unbounded gapless 2D graphene monolayers.
However, the opened band gap in GNR reduces considerably the mobility
of the carriers \cite{Wang13}. Additional inevitable difficulties come
from the fact that in contrast to gapless graphene monolayers, in which the bound
impurity states are forbidden \cite{Bis,Nov,Per,Shyt,Shyt1} in gapped graphene
\cite{Nov,Peder,Gupta08} and quasi 1D GNR \cite{monschm12} the bound impurity
states can be realized. The binding energies $E^{(b)}$ of the impurity electrons
in the GNR of width $1~\mbox{nm}$ reach the considerable amount of the order
of $E^{(b)}\simeq 1\mbox~{eV}$. In addition, it was shown \cite{monschm12} that impurity electrons,
possessing energies close to the Breit-Wigner meta-stable resonances, contribute
negligibly little to the conductance. Clearly, the impurity centres suppress
strongly the mobility of the GNRs. Of course
the binding effect of the impurity centres could be reduced by the technologically
involved procedure of the improvent
of the sample standard \cite{Wang13}.
Nevertheless the less elaborate mechanism of the liberation of the carriers
captured by the impurities is a much more immediate demand at the present time. The process
of the ionization by an external electric field can be used as an
instrument for the release of the blocked carriers.

Besides, the quasi-1D structures, in particular the bulk semiconductors subject
to strong magnetic fields \cite{ZhMak}, quantum wires (QWRs) \cite{monschm09}
and armchair GNRs \cite{monschm12} are favorable for the formation of both
the strictly discrete and meta-stable (Fano resonances) \cite{Fano}
impurity and exciton states adjacent to the ground and excited
size-quantized energy levels, respectively. The latter are caused
by the confinement effect associated with the magnetic field
in bulk semiconductors and the boundaries
of the QWRs and GNRs. The nature of the Fano resonances comes from the
inter subband coupling between the discrete and low-lying continuous Rydberg
states. With emergence of the electric field only one channel of the ionization
is opened for the ground series of the Rydberg discrete states, while the
excited meta-stable states decay into two channels: the autoionization channel,
open due to the inter-subband Fano coupling, and the channel of the electric
field ionization, related to the under-barrier tunneling (Fig. 1).
The interaction of these two channels of the ionization is of immediate interest.

\begin{figure}[htbp]
\begin{center}
\includegraphics[width=8.6cm]{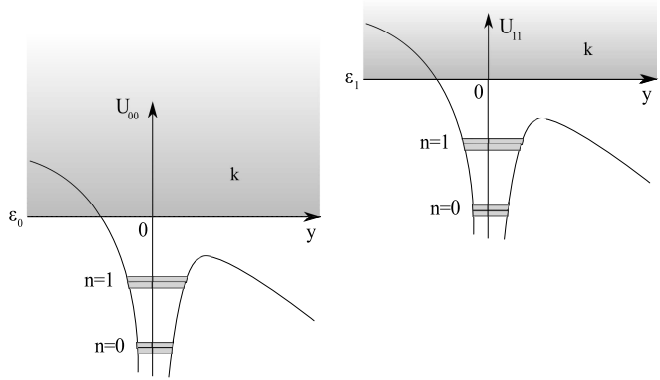}
\end{center}
\caption{\label{qd1} (color online). The combined potential $U_{NN}(y)=V_{NN}(y)-eFy$, formed by the impurity
potential $V_{NN}(y)$ (\ref{E:pot}) and electric field $F$ adjacent to the
ground and first excited size-quantized energy levels
$\varepsilon_{N}$ (\ref{E:ener}) with $N=0,1$, respectively. The continuous and discrete
Rydberg states are labeled by the indexes $(k)$ and $n=0$ (ground state) and
$n=1$ (first excited state), respectively.}
\end{figure}

It is clear that the problem of the impurity states in the armchair GNR
in the presence of a longitudinal electric field
directed parallel to the GNR axis is important on account
of two aspects: (i) its considerable interest in the context of basic research,
and (ii) possible
nano-electronic applications.

Narrow GNRs of several
nanometers width are the best candidates among the other
1D structures for fundamental studies.
The binding energies of the impurity electron in GNR exceed
those in the corresponding semiconductor structures by a factor of about $10^2$
that in particular manifests itself in the strong electric fields,
providing the complete ionization of the impurity states in GNR. GNRs seem to be
a unique structure in which both
channels of the auto- and electric field ionization are opened. The process
of electric field ionization transforms the strictly bound and the Fano resonant states
into states of transporting carriers that in turn improve the conductance properties
of the GNRs and of the nano-electronic devices into which these ribbons are
incorporated. The finite lifetime of the quasi-discrete impurity states
associated with the two-channel ionization should be taken into account
in the practical use of the GNRs exposed to the electric field.

There are two comments in order. First, the theoretical
approaches to this problem are mostly based on numerical calculations
(density functional theory and Bethe-Salpiter equation \cite{Yang07},
nonorthogonal tight-binding model \cite{Jia}, tight-binding scheme and
Hartree-Fock approximation \cite{Mohamm}) requiring significant computational
efforts. Only a few recent works elaborate on analytical methods.
In Ref. \cite{monschm12} the bound and quasi-discrete impurity states in
the armchair GNR have been studied by solving the Dirac equation for a
massless neutrino. Ratnikov and Silin \cite{RatSil} empirically extended
to the GNR the model earlier developed for the semiconductor QWR \cite{Bab},
and calculated the excitonic energy levels by the variational method
and their red shift induced by the electric field. Second, to our knowledge
analytical results based on the Dirac equation adequately describing
the impurity electrons in GNRs subject to external electric field are not
present in the literature. Thus an analytical approach to the problem
of the impurity states in biased armchair GNR is desirable.
Particularly it renders the basic physics transparent and governs the electronic,
optical and transport properties
of the graphene based devices.

In the present work we develop an analytical approach to the problem of the
impurity state in the narrow armchair GNR in the presence of an external
electric field directed parallel to the ribbon axis. The Coulomb impurity
attraction is taken to be much weaker than the influence of the ribbon
confinement and much stronger than the effect of the electric field.
The impurity centre can be positioned anywhere within the GNR. The 2D Dirac
equation for the massless neutrino subject to the Coulomb and
external uniform electric field
is solved in the adiabatic approximation. This approximation implies
the transverse motion
of the electron governed by the ribbon confinement to be much faster than
the longitudinal motion controlled by the impurity and external
electric field. Our approach is based on the matching
of the
wave functions in the intermediate regions. The latter
separates the impurity interaction from the electric field interaction
dominated regimes.
In the approximation of the isolated size-quantized  subbands the binding
energies and widths of the quasi-discrete states as a function
of the ribbon width, position of the impurity and the electric field are
calculated in explicit form. Also, the phases of the functions of the
continuous spectrum specifying the S-matrix are derived.
In the double-subband approximation the total widths of the first excited
Rydberg series of impurity states, associated with the ionization effect of the electric field
and inter-subband Fano coupling are calculated. Also the capturing of the electron
by the impurity potential for the lifetime determined by the electric field
is explored.
Numerical estimates made for realistic GNR show
that for narrow ribbons the impurity states in the presence of a weak electric
field remain quite stable which is to be proven experimentally, while significantly
strong field could unlock
the captured electrons. The aim of this work is to clarify
the ionization mechanism of the release of the strictly bound and
quasi-bound impurity electrons yielding the increase of the mobility
of the carriers in the GNR.

This work is organized as follows. In Section 2 the general analytical
approach is described. The complex quasi-discrete energy levels dictating
the binding energies and energy widths caused by the electric field
along with the phases of the wave functions of the continuous states
are calculated in the
single-subband approximation in Section 3. The combined effect of the
autoionization of the Fano resonant states and their ionization by the
electric field is under consideration in Section 4. In Section 5 we discuss
the obtained results and estimate
the expected experimental values. Section 6 contains the conclusions.

\section{General approach}\label{S:gen}

We consider a ribbon of width $d$ placed in the $x-y$ plane and bounded by the lines $x=\pm d/2.$
The impurity centre of charge $Z$ is shifted from the mid-point of the ribbon
$x=0$ by the distance $-d/2 \leq x_0 \leq d/2.$
The equation describing the impurity electron at a position $\vec{\rho}=(x,y)$
subject to the external uniform electric field $\vec{F}\parallel -\vec{e}_y$
possesses the form of a Dirac equation

\begin{equation}\label{E:basic}
\hat{{\rm H}}(\hat{\vec{k}},\vec{\rho })\vec{\Psi}(\vec{\rho })=E\vec{\Psi}(\vec{\rho });\qquad \hat{\vec{k}}=-i\vec{\nabla};
\end{equation}
where the Hamiltonian $\hat{{\rm H}}$ relevant to the inequivalent Dirac points\\
$\vec{K}^{(+,-)}(\pm K, 0)~;\, (K=4\pi/3a_0,~a_0 = 2.46 ~{\rm {\AA}}$
is the graphene lattice constant) is given by \cite{monschm12,Brey}

$$
\hat{{\rm H}}= p\left[ \left(\begin{array}{cc} -\sigma_x \hat{k}_x
& 0 \\0 & \sigma_x \hat{k}_x \\
\end{array} \right) + \left (\begin{array}{cc} -\sigma_y
\hat{k}_y & 0 \\0 & - \sigma_y \hat{k}_y \end{array} \right)      \right]
 +\left[V(\vec{\rho}) - eFy\right]\hat{{\rm I}};
$$
containing the Pauli matrixes $\sigma_{x,y}$, the graphene parameter
$p=\hbar v_F ;\,(v_F=10^6~\mbox{m/s})$, the unit matrix  $\hat{{\rm I}}$ and
the 2D Coulomb impurity potential

\begin{equation}\label{E:coulomb}
V(\vec{\rho})=-\frac{\beta}{\sqrt{(x-x_0)^2 + y^2}}~;\qquad \beta = \frac{Ze^2}{4\pi\epsilon_0 \epsilon_{\mbox{eff}}}~;
\end{equation}
Here $\epsilon_{\mbox{eff}} =\frac{1}{2}(1+\epsilon +\pi q_0)$
is the effective dielectric constant determined by the static dielectric constant
$\epsilon$ of the substrate  \cite{Nov,hwang} and by the
parameter $q_0 = \frac{e^2}{4\pi\epsilon_0\hbar v_F}\simeq2.2.$

The envelope wave function four-vector $\vec{\Psi}$

$$
\vec{\Psi}(\vec{\rho }) =
\begin{array}{c}
\begin{Bmatrix}
\psi_{A}^{(+)} \\
0 \\
\psi_{A}^{(-)}\\
0
\end{Bmatrix}
\end{array}
+
\begin{array}{c}
\begin{Bmatrix}
 0\\
\psi_{B}^{(+)} \\
0 \\
\psi_{B}^{(-)} \\
\end{Bmatrix}
\end{array}
$$
consists of the wave functions $\psi^{(+,-)}_{A,B}(\vec{\rho})$
describing the electron states in the sublattices $A$ and $B$ of graphene
in the vicinity of the Dirac points $\vec{K}^{(+,-)}$, respectively.
The boundary conditions require the total wave function to vanish
at both edges for each sublattice \cite{Castro}

\begin{equation}\label{E:bound}
 e^{{\rm i}Kx}\psi_{j}^{(+)}(\vec{\rho})+ e^{-{\rm i}Kx}\psi_{j}^{(-)}(\vec{\rho})=0
  ~\mbox{at}\, x=\pm\frac{d}{2}, ~\mbox{for}~j = \mbox{A,B}
\end{equation}
By solving eq. (\ref{E:basic}) the components $\psi^{(+,-)}_{A,B}$ of the
total wave vector $\vec{\Psi}(\vec{\rho})$ subject to the boundary
conditions (\ref{E:bound}) can be found.

Following the procedure presented in details in Ref. \cite{monschm12}
we expand the wave functions $\psi^{(+,-)}_{A,B}$ in a series

$$
\psi_{j}^{(+,-)}(\vec{\rho})=\Sigma_N u_{Nj}(y)\varphi_{Nj}^{(+,-)}(x),~j = \mbox{A,B},
$$
in which

\begin{eqnarray}
-\varphi_{NA}^{(+)}&=&\varphi_{NA}^{(-)*}=\varphi_{NB}^{(+)}=-\varphi_{NB}^{(-)*}=\varphi_{N0};
\nonumber\\
\varphi_{N0}(x)&=&\frac{1}{\sqrt{2d}}\exp \left \{ {\rm i}\left [x\frac{\pi}{d}(N-\tilde{\sigma})-\frac{\pi}{2}
\left(N+ \left[ \frac{Kd}{\pi} \right]    \right)  \right ]\right \}
\end{eqnarray}
are the components of the orthonormal $N$-vector wave function relevant to the transverse confined $x$-motion
of the free electron with the size-quantized energies

\begin{equation}\label{E:ener}
\varepsilon_N=|N-\tilde{\sigma}|\frac{\pi p}{d};~N=0,\pm1,\pm2,\ldots~;
\quad\tilde{\sigma}=\frac{Kd}{\pi}-\left[ \frac{Kd}{\pi}\right].
\end{equation}
Below for estimates we take the GNRs of the family $\tilde{\sigma}=1/3$
providing along with $\tilde{\sigma}=2/3$ the semiconductor-like gapped
structure, leaving aside $\tilde{\sigma}=0$, corresponding to the
metallic-like gapless ribbon. This
leads to the set of the equations for the coefficients

$$
v_N^{(1,2)}=\frac{1}{\sqrt{2}}(u_{NB}\pm u_{NA});
$$

\begin{eqnarray}\label{E:set}
\left.
\begin{array}{c}
\frac{dv_N^{(1)}(y)}{dy} - \frac{1}{p}\left(E + \varepsilon_N -V_{NN}(y)+ eFy\right)v_N^{(2)}(y)
+\frac{1}{p}\sum_{N'\neq N} V_{N'N}(y)v_{N'}^{(2)}(y)=0~;\\
\frac{dv_N^{(2)}(y)}{dy} + \frac{1}{p}\left(E - \varepsilon_N -V_{NN}(y)+eFy\right)v_N^{(1)}(y)
-\frac{1}{p}\sum_{N'\neq N} V_{N'N}(y)v_{N'}^{(1)}(y)=0~;
\end{array}
\right \}
\end{eqnarray}

\begin{equation}\label{E:pot}
V_{N'N}(y)=\frac{1}{d}\int_{-\frac{d}{2}}^{+\frac{d}{2}} V(\vec{\rho})
\cos \left[(N-N')\pi \left(\frac{x}{d}- \frac{1}{2}\right)dx  \right],
\end{equation}
with eq. (\ref{E:coulomb}) for the potential $V(\vec{\rho})$.
At $|y|\gg d$

\begin{equation}\label{E:onoff}
V_{N'N}(y)=-\frac{\beta}{|y|}\left[ \delta_{N'N}+O \left(\frac{d^2}{y^2}  \right)
\delta_{|N'-N|(2s+1)} \right]~;\,s=0,1,2,\ldots ;
\end{equation}

Below we solve the set (\ref{E:set}) in the adiabatic approximation.
The longitudinal $y$-motion, governed by the quasi-Coulomb potentials
$V_{N'N}(y)$ slightly perturbed by the electric field $F$, is assumed
to be much slower than the transverse $x$-motion affected by the boundaries
of the narrow ribbon.

The relevant parameters are the strength of the impurity
potential scaled to that of the graphene $q$, the impurity Bohr radius
$a_0$, the quantum number of the bound impurity state $\eta$ and the
dimensionless electric field $f$, which is the external electric field $F$
scaled to the impurity electric field $F_0$. They are defined by

\begin{equation}\label{E:par}
q=\frac{\beta}{p};\quad a_0 = \frac{p}{E q};\quad
\eta = q\frac{E}{\sqrt{\varepsilon_N^2 - E^2}};\quad f=\frac{F}{F_0};\quad
F_0=\frac{qp}{e a_0^2};
\end{equation}
The other parameters $y_1 =2\eta^2 a_0$ and $y_2 = (\varepsilon_N -E)(eF)^{-1}$ -
are the first and second quasi-classical turning points calculated from
$\mathcal{P}(y_{1,2})=0$, where

$$
\mathcal{P}(y) =
\frac{1}{v_F} \left[\left( E +\frac{\beta}{y} +eFy \right)^2 - \varepsilon_N^2 \right]^{1/2};
$$
is the quasi-classical momentum. Further we impose the conditions

\begin{equation}\label{E:weakimp}
q\ll 1
\end{equation}
meaning the narrowness of the ribbon $d \ll a_0$
(at any rate for the low excited size-quantized $N$ subbands)
i.e. the smallness of the impurity effect
comparatively to that of the confinement, and

\begin{equation}\label{E:weakel}
f\eta^3 \ll 1
\end{equation}
providing the weakness of the external electric field $F$ relatively to the impurity
electric field in the state with quantum number $\eta$. Under these conditions
the relationships

$$
a_0, y_1 \ll y_2,\, a_{0}= \frac{d}{\pi |N-\tilde{\sigma}| q},~y_2 = \frac{\varepsilon_N q^2}{2\eta^2 eF}
$$
are valid.

\section{Single-subband approximation}\label{S:single}

At the first stage we neglect the coupling between the states associated with the
subbands of different $N.$ It follows from eq. (\ref{E:onoff}),
that in the narrow ribbon of small width $d$ the diagonal potentials $V_{NN}$
dominate the off-diagonal terms which allows in turn allows to take
$V_{N'N}=V_{NN}\delta_{N'N}$ and then to decompose the set (\ref{E:set}) into
independent equations with the potentials

\begin{eqnarray}\label{E:limit}
V_{NN}(y)=\frac{\beta}{d}\ln\frac{\frac{4y^2}{d_1 d_2}}{\left(1 + \sqrt{1+\frac{4y^2}{d_1 ^2}} \right)
\left(1 + \sqrt{1+\frac{4y^2}{d_2
^2}} \right)}=
\left\{
\begin{array}{cl}
\frac{\beta}{d}\ln\frac{y^2}{d_1 d_2}~;\, &\frac{|y|}{d_{1,2}}\ll 1\\
-\frac{\beta}{|y|}~;\, &\frac{|y|}{d_{1,2}}\gg 1
\end{array}
\right.
\end{eqnarray}

$$
d_{1,2}=d\pm 2x_0~;\qquad -\frac{d}{2} \leq x_0 \leq +\frac{d}{2}~;
$$

The set (\ref{E:set}) for $V_{N'N}=0$
is solved by matching in the intermediate regions
the two-vectors $\vec{V}_N =(v_N^{(1)},v_N^{(2)})$ valid in the inner region
$0\leq y \ll a_0$, Coulomb region $d \ll y \ll y_2$ and in the "electric" region $a_0 \ll y$
\cite{monschm12}. In the inner and Coulomb regions the impurity electric field $F_0$
considerably exceeds the external uniform field $F$, while in the "electric" region
the potentials $V_{NN}(y)$ can be treated as a small perturbation to the effect
of the field $F$.

\subsection{Discrete states $E < \varepsilon_N$}

\emph{Inner region}
\\
\\
In this region
an iteration procedure is employed. The subsequent integration
of the set (\ref{E:set}), in which we keep only diagonal potentials $V_{NN}(y)$
(\ref{E:limit}) and take arbitrary constants for the trial functions $v_N^{(1,2)}$,
gives for the even states $\vec{V}_N$ in the intermediate region $d \ll y \ll a_0$
\cite{monschm12}

\begin{equation}\label{E:iter}
v_{N \mbox{it} }^{(1)}(y)=\sin(Q(y)+\zeta);\quad v_{N \mbox{it}}^{(2)}(y)=\cos(Q(y)+\zeta)~;
\end{equation}
where

$$
Q(y)=q\frac{y}{|y|}\left( \ln\frac{4|y|}{D}+1 \right),\quad
D=\sqrt{d_1 d_2}\exp \left\{ \frac{1}{4d}(d_1 - d_2)\ln\frac{d_1}{d_2}\right\}
$$
and $\zeta$ is an arbitrary constant phase.
\\
\\
\emph{Coulomb region}
\\
\\
In this region the wave two-vector $\vec{V}_{N\mbox{C}}$ can be written in the form

\begin{equation}\label{E:coulvec}
\vec{V}_{NC}(y)=R_{-}\vec{V}_{NC}^{(-)}(y) + R_{+}\vec{V}_{NC}^{(+)}(y)
\end{equation}
where $\vec{V}_{NC}^{(+)}~\mbox{and}~\vec{V}_{NC}^{(-)}$ are the vectors
increasing and decreasing, respectively at $|y|\rightarrow \infty$, and
where $R_{\pm}$ are the corresponding arbitrary constants. The components determining
the vector $\vec{V}_{NC}^{(-)}$, have been calculated in Ref. \cite{monschm12}
in terms of the exact solutions to eqs. (\ref{E:set}) at
$V_{NN}(y)=-\beta |y|^{-1}~\mbox{and}~F=0$

\begin{equation}\label{E:outer}
v_{NC(-)}^{(1)}(y)=\cosh\frac{\psi}{2}\tau^{-\frac{1}{2}}\left[ W_{\kappa,\mu}(\tau)
+\frac{\tanh \psi}{q}W_{\kappa +1,\mu}(\tau) \right],
\end{equation}
where

$$
\tau = \frac{2}{\eta a_0}y;~\quad\tanh \psi = \frac{q}{\eta};\quad\kappa = \eta -\frac{1}{2};
\quad\mu = {\rm i }q,
$$
and where $W_{\kappa,\mu}(\tau)$ is the Whittaker function having the asymptotics
$\exp (-\frac{\tau}{2})$ \cite{abram}.
The function $v_{NC(-)}^{(2)}(y)$ can be obtained from eq. (\ref{E:outer})
by replacing $\cosh\frac{\psi}{2}~\mbox{by}~\sinh\frac{\psi}{2}~\mbox{and}~q~\mbox{by}~-q$.
The wave functions $v_{NC(+)}^{(1,2)}$, corresponding to the vector $\vec{V}_{NC}^{(+)}$,
are derived from the functions $v_{NC(-)}^{(1,2)}$, respectively by replacing
$W_{\kappa,\mu}~\mbox{by}~M_{\kappa,\mu},~W_{\kappa +1,\mu}~\mbox{by}~M_{\kappa +1,\mu}
~\mbox{and}~q~\mbox{by}~q(\eta + {\rm i }q)^{-1}~\mbox{where}~M_{\kappa,\mu}(\tau)$ is the
Whittaker function having the asymptotics $\exp (\frac{\tau}{2})$ \cite{abram}.

At $\tau \ll 1~\mbox{and}~q\ll 1$

\begin{equation}\label{E:outer1}
v_{NC(-)}^{(1)}(y)=-\frac{1}{\eta^2 \Gamma(-\eta)}\sin\omega (y);\quad
v_{NC(-)}^{(2)}(y)=-\frac{1}{\eta^2 \Gamma(-\eta)}\cos\omega (y);
\end{equation}

\begin{equation}\label{E:outer2}
v_{NC(+)}^{(1)}(y)=\frac{q}{\eta}\sin (q\ln\tau);\quad
v_{NC(+)}^{(2)}(y)=\frac{q}{\eta}\cos (q\ln\tau);
\end{equation}
where $\omega (y)=q\ln\tau +\Theta (\eta)$ with

\begin{equation}\label{E:theta}
\Theta (\eta)=\frac{\pi}{2}+2qC+\arg \Gamma(-\eta + {\rm i}q)-\frac{q}{2\eta}.
\end{equation}
In eq. (\ref{E:theta}) $C=0.577$ is the Euler constant and $\Gamma(x)$
is the $\Gamma$-function.

At $\tau \gg 1$

\begin{equation}\label{E:outer3}
v_{NC(-)}^{(1)}(y)=\frac{1}{\eta}\tau^{\eta}{\rm e}^{-\frac{\tau}{2}};\quad
v_{NC(-)}^{(2)}(y)=-\frac{q}{2\eta}v_{NC(-)}^{(1)}(y);
\end{equation}

\begin{eqnarray}\label{E:outer4}
\left.
\begin{array}{c}
v_{NC(+)}^{(1)}(y)=-\frac{1}{\eta}
\left [\frac{{\rm e}^{-\rm i \pi\eta}}{\Gamma (\eta)}\tau^{\eta}{\rm e}^{-\frac{\tau}{2}}+
\frac{1}{\Gamma (-\eta)}\tau^{-\eta}{\rm e}^{+\frac{\tau}{2}}   \right];\\
v_{NC(+)}^{(2)}(y)=\frac{q}{2\eta^2}
\left [\frac{{\rm e}^{-\rm i\pi\eta}}{\Gamma (\eta)}\tau^{\eta}{\rm e}^{-\frac{\tau}{2}}-
\frac{1}{\Gamma (-\eta)}\tau^{-\eta}{\rm e}^{+\frac{\tau}{2}}   \right];
\end{array}
\right \}
\end{eqnarray}

\emph{"Electric region"}
\\
\\
The problem of the relativistic electron in the presence of a
uniform electric field has been studied initiatively by Sauter \cite{Sauter}.
Using the original notations

$$
  \xi = \sqrt{\frac{eF}{p}}y + \xi_0;\quad\xi_0=\frac{E}{\sqrt{eFp}};\quad
  k_0=\frac{\varepsilon_N}{\sqrt{eFp}};
$$
the set (\ref{E:set}) for the functions $v_{N\mbox{el}}^{(1,2)}$ reads

\begin{eqnarray}\label{E:setel}                                           \left.
\begin{array}{c}
v_{N\mbox{el}}^{(1)''}(\xi) +
 \frac{-1 + \frac{q}{(\xi-\xi_0)^2}}{k_0 + \xi +\frac{q}{\xi-\xi_0}}v_{N\mbox{el}}^{(1)'}(\xi)-
 \left [k_0^2 - \left (\xi + \frac{q}{\xi-\xi_0} \right)^2 \right ] v_{N\mbox{el}}^{(1)}(\xi)=0 ~;\\
v_{N\mbox{el}}^{(2)}(\xi) =
 \frac{1 }{k_0 + \xi +\frac{q}{\xi-\xi_0}}v_{N\mbox{el}}^{(1)'}(\xi)~;
\end{array}
\right \}
\end{eqnarray}

Using the relationships $k_0,~\xi_0,~\xi \gg 1$ induced by the conditions
(\ref{E:weakimp}) and (\ref{E:weakel}) and setting

\begin{eqnarray}\label{E:subst}
&\ & v^{(1)}_{N\mbox{el}}(\xi) = (\xi + k_0)^{\frac{1}{2}}\varphi(\xi)(x)~;
\nonumber
\\
&\ & \xi = k_0 \left [1- (2k_0^4)^{-\frac{1}{3}}x     \right ]~,
x=(2k_0)^{\frac{1}{3}}\sqrt{\frac{eF}{p}}(y - y_2);\quad x\ll k_0^{\frac{4}{3}};
\end{eqnarray}
we obtain from eq. (\ref{E:setel})

\begin{equation}\label{E:gfunct}
\varphi^{''}(x) - G(x)\varphi(x)=0~,
\end{equation}
where

$$
G(x)=x-2q(2k_0)^{-\frac{2}{3}} \frac{\xi_0}{k_0 - \xi_0 - (2k_0)^{-\frac{1}{3}}x }.
$$

Eq. (\ref{E:gfunct}) is solved by the method of a comparison equation
\cite{Slav} successfully employed in Ref. \cite{monschm09} in which the
impurity and exciton
in a biased quantum wire have been studied. The key point of this method
is the replacements of the coefficient $G(x)$ and the function $\varphi (x)$
by others which transform eq. (\ref{E:gfunct}) into an exactly analytically solvable
comparison equation (see Refs. \cite{Slav} and \cite{monschm09} for details).
The solutions to eq. (\ref{E:gfunct}) $\varphi_{1,2}(x)$ are written
in terms of the Airy functions $Ai(S)$ and $Bi(S)$ \cite{abram}

\begin{equation}\label{E:Airy}
\varphi_1 (x) = \frac{\left[\frac{3}{2}S(x)  \right]^{\frac{1}{6}}}{G(x)^{\frac{1}{4}}}
Ai\left[\left(\frac{3}{2}S(x)  \right)^{\frac{2}{3}} \right]~;\quad
\varphi_2 (x) = \frac{\left[\frac{3}{2}S(x)  \right]^{\frac{1}{6}}}{G(x)^{\frac{1}{4}}}
Bi\left[\left(\frac{3}{2}S(x)  \right)^{\frac{2}{3}} \right]~;
\end{equation}
where

\begin{equation}\label{E:momel}
S(x)= \int_0 ^x G^{\frac{1}{2}}(x)dx.
\end{equation}

At $y \ll y_2$ resulting in $x, S \gg 1$, the asymptotic expansions
for $Ai(S)~\mbox{and}~Bi(S)$ \cite{abram} in eqs. (\ref{E:Airy})
give for the functions $v_{N\mbox{el}}^{(1)}$ (\ref{E:subst}) and
$v_{N\mbox{el}}^{(2)}$ (\ref{E:setel})

\begin{eqnarray}\label{E:elcomp}
v_{N\mbox{el}}^{(1)(-)}(y)& = & \pi^{-\frac{1}{2}}x_2^{-\frac{1}{4}}\exp \left[  S(y) \right]~;~
v_{N\mbox{el}}^{(1)(+)}(y) = \pi^{-\frac{1}{2}}x_2^{-\frac{1}{4}}\frac{1}{2}\exp \left[ - S(y) \right]~;
\nonumber
\\
v_{N\mbox{el}}^{(2)(\mp)}(y)& = & \mp \frac{q^2}{4\eta^2}v_{N\mbox{el}}^{(1)(\mp)}(y)~;
\quad x_2 = (2k_0)^{\frac{1}{3}}\sqrt{\frac{eF}{p}}y_{2}~;
\end{eqnarray}
where

\begin{equation}\label{E:momas}
S(y)=\frac{1}{3\eta^3 f} - \frac{y}{\eta a_0} + \eta \ln \frac{y}{4y_2}.
\end{equation}

The components $v_{N\mbox{el}}^{(1,2)(\pm)}$ in eqs. (\ref{E:elcomp}) determine
the two-vector $\vec{V}_{N\mbox{el}}$ in the region $y\ll y_2$

\begin{equation}\label{E:genel}
\vec{V}_{N\mbox{el}}(y) = C_0\left[\vec{V}_{N\mbox{el}}^{(-)}(y)+
{\rm i} \vec{V}_{N\mbox{el}}^{(+)}(y)  \right],
\end{equation}
where $C_0$ is an arbitrary constant. Note, that in the region
$y > y_2$ the vector state (\ref{E:genel}) with eqs.
(\ref{E:setel}), (\ref{E:subst}),
(\ref{E:Airy}), possesses the asymptotics of the outgoing wave

$$
\vec{V}_{N\mbox{el}}(y)=C_0 \pi^{-\frac{1}{2}}x^{-\frac{1}{4}} \exp \left\{{\rm i}\left[
\frac{k_0^2}{3}\left(\frac{2eF}{\varepsilon_N} (y-y_2)  \right)^{\frac{3}{2}}
 +\frac{\pi}{4}   \right]      \right\}.
$$

On equating in the intermediate region $d \ll y \ll a_0$ the two-vectors $\vec{V}_{N\mbox{it}}$
and $\vec{V}_{NC}$ (\ref{E:coulvec}) with the components (\ref{E:iter}) and
 (\ref{E:outer1}), (\ref{E:outer2}) for the vectors $\vec{V}_{N\mbox{it}}$ and
$\vec{V}_{NC}$, respectively, we obtain

\begin{equation}\label{E:eq1}
-R_{-}\frac{1}{\Gamma(-\eta +1)} Y(\eta) + R_{+}=0
\end{equation}
with

\begin{equation}\label{E:y}
Y(\eta)=\frac{1}{q}\left[\Theta (\eta) -q \left( \ln\frac{2\eta a_0}{D} + 1 \right)
- \zeta  \right].
\end{equation}
Taking in eq.(\ref{E:theta}) $\Theta (\eta)~\mbox{for}~\zeta =\pi/2 ~\mbox{and}~q\ll 1$,
the function $Y(\eta)$ reads in an explicit form

\begin{eqnarray}\label{E:Y}
Y &=& q^{-1}\left[\arctan \left(\frac{1}{z} - \frac{n}{q}  \right)^{-1} - \arctan\frac{z}{2} \right]
+ \ln z + \psi\left( 1+ \frac{q}{z}\right) + X_N~;
\\
 X_N &=& \ln \left\{\frac{|N - \tilde{\sigma}|\pi }{2}\sqrt{1-s^2}\exp\left [\frac{s}{2}\ln\frac{1+s}{1-s} \right ]\right\}
+ 2C - 1~;
\nonumber
\end{eqnarray}
In eq. (\ref{E:Y}) $z = q/\eta~, s=2x_{0}/d~,\quad \eta = n + \delta_{Nn}~,\quad n = 0,1,2\ldots ,$
and $\psi(x) = \Gamma' (x)/\Gamma (x)$ is the logarithmic derivative of the $\Gamma (x)$-function.
In an effort to make the further results more readable and transparent, we utilize
the logarithmic approximation $|q\ln q|\ll 1~(z \ll 1)$~, which transforms eq. (\ref{E:Y}) into

\begin{equation}\label{E:Y1}
Y(\eta) = \frac{1}{\eta - n} - \frac{1}{2\eta } + \ln q - \ln \eta + \psi\left( 1+ \eta \right) + X_N~;~
n = 0,1,2\ldots
\end{equation}
for the $Y$ function and for its derivative we obtain

\begin{equation}\label{E:der}
\frac{\partial Y}{\partial \eta} =  - \frac{1}{(\eta - n)^2}+\frac{1}{2\eta^2}~,\quad n = 0,1,2\ldots .
\end{equation}

A comparison in the other intermediate region $a_0 \ll y \ll y_2$ the Coulomb vector
$\vec{V}_{NC}$ (\ref{E:coulvec}), (\ref{E:outer1}), (\ref{E:outer2})
and "electric" vector $\vec{V}_{N\mbox{el}}$ (\ref{E:genel}), (\ref{E:elcomp})
yield

\begin{eqnarray}\label{E:eqel}
\left.
\begin{array}{c}
-R_{-}+ R_{+}\frac{(-1)^{\eta}}{\Gamma(\eta )} + C_0\pi^{-\frac{1}{2}}x_2^{-\frac{1}{4}}\eta \Phi_{\eta}^{-1}=0~;\\
R_{+}\frac{1}{\Gamma(-\eta )} + C_0 \pi^{-\frac{1}{2}}x_2^{-\frac{1}{4}}{\rm i}\frac{\eta}{2} \Phi_{\eta}=0~;
\end{array}
\right \}
\end{eqnarray}
where

\begin{equation}\label{E:Phi}
\Phi_{\eta} = \exp \left(-\frac{1}{3\eta^3 f} +\eta \ln\frac{4}{\eta^3 f} \right)~,
\Phi_{\eta} \ll 1.
\end{equation}

On solving the set (\ref{E:eq1}), (\ref{E:eqel}) by the determinantal method
we arrive at the equation for the complex quantum numbers $\eta$ and complex
energies $E$

\begin{equation}\label{E:compleq}
-\frac{1}{2}\Gamma(1-\eta)\left[\frac{\Gamma(-\eta)}{Y(\eta)}
 +\frac{(-1)^{\eta}}{\Gamma(1+\eta)}   \right]\Phi_{\eta}^2 +{\rm i} = 0.
\end{equation}

The quantum numbers $\eta_{Nn} = n + \delta_{Nn}~, n = 0,1,2\ldots $
of the strictly discrete states related to the zero electric field $F=\Phi_{\eta}=0 $
can be found from equation $Y(\eta)=0$ with eq. (\ref{E:Y1}) for the $Y$ function.
On expanding this function in eq. (\ref{E:compleq}) in the vicinity of the
quantum numbers $\eta_{Nn}$ and taking into account the derivative (\ref{E:der}) we calculate
the complex quantum numbers $\eta(E)$ which in turn determine the quasi-discrete energy levels

\begin{equation}\label{E:elen}
E_{Nn}= \varepsilon_{N}\left(1 - \frac{q^2}{2\eta^2}   \right) - {\rm i}\frac{\Gamma_{Nn}^{(\mbox{el})}}{2}~;
\end{equation}
where the energy width

\begin{eqnarray}\label{E:width}
\Gamma_{Nn}^{(\mbox{el})}=
\left\{
\begin{array}{cl}
2\delta_{N0}^{-2}\varepsilon_{N}q^2\Phi_0^2~;\quad n=0~;
\\
\\
(nn!)^{-2}\varepsilon_{N}q^2\Phi_n^2~;\quad n=1,2,\ldots~;
\end{array}
\right.
\end{eqnarray}

Replacing the vector (\ref{E:genel}) by the "electric" vector $\vec{V}_{N\mbox{el}}$

\begin{equation}\label{E:stand}
\vec{V}_{N\mbox{el}} = C_0\left[\cos\Omega_0\vec{V}_{N\mbox{el}}^{(-)}(y) -
\sin\Omega_0\vec{V}_{N\mbox{el}}^{(+)}(y)  \right],
\end{equation}
where $C_0$ and $\Omega_0$ are the arbitrary constant and phase, respectively,
we obtain

\begin{equation}\label{E:cot}
\cot\Omega_0 = \left \{-\frac{1}{2}\left[\frac{\Gamma(-\eta)}{Y} +
 \frac{(-1)^{\eta}}{\Gamma(1+\eta)} \right]\Gamma(1-\eta)\Phi_{\eta}^2 \right\}^{-1}
\end{equation}
The results of this subsection summarized in eqs. (\ref{E:elen}), (\ref{E:width}),
and (\ref{E:cot}) are valid under the conditions
(\ref{E:weakimp}) as well as $|q\ln q| \ll 1$ and (\ref{E:weakel}).

\subsection{Continuous states $E > \varepsilon_N$}

\emph{Inner region}
\\
\\
As above the wave functions, corresponding to the inner region are given by eqs. (\ref{E:iter}).
\\
\\
\emph{Coulomb region}
\\
\\
In the region $d \ll y \ll \tilde{y}_2$, where
 $\tilde{y}_2 = (\xi_0 - k_0)p^{\frac{1}{2}}(eF)^{-1/2}$, the two-vector
 $\vec{V}_{NC}$ reads

\begin{equation}\label{E:coulcont}
\vec{V}_{NC}(y)={\rm e}^{{\rm i}\Omega}\vec{V}_{NC}^{(+)}(y) +
{\rm e}^{-{\rm i}\Omega}\vec{V}_{NC}^{(-)}(y),
\end{equation}
where $\Omega$ is an arbitrary phase. The arbitrary constants analogous to those
in eqs. (\ref{E:genel}) and (\ref{E:stand}) do not contribute to the results
of this paragraph and are therefore omitted.
The
components of the vectors $\vec{V}_{NC}^{(+)}=\vec{V}_{NC}^{(-)*}$
were calculated in Ref. \cite{monschm12} in terms of the exact solutions
to eqs. (\ref{E:set}) at $V_{NN}(y)= -\beta |y|^{-1}$ and $F=0$. In particular

\begin{equation}\label{E:outcont}
v_{NC(+)}^{(1)}=\cos\frac{\varphi_N}{2}t^{-\frac{1}{2}}\left[ W_{\tilde{\kappa},\mu}(t)
- {\rm i}\frac{\tan \varphi_N}{q}W_{\tilde{\kappa}+1,\mu}(t) \right],
\end{equation}
where

$$
t=-2{\rm i}ky;~k = \frac{1}{p}\sqrt{E^2 - \varepsilon_N^2};\,\tan\varphi_N =\frac{pk}{\varepsilon_N};\,
\tilde{\kappa} = {\rm i}\frac{q}{\sin\varphi_N}-\frac{1}{2};
$$
The function $v_{NC(+)}^{(2)}$ can be obtained from eq. (\ref{E:outcont})
 by replacing $\cos\frac{\varphi_N}{2}$ by $-{\rm i}\sin\frac{\varphi_N}{2}$
 and $q$ by $-q$.

For $ky \ll 1$ the components $v_{NC}^{(1,2)}$ of the vector $\vec{V}_{NC}$
(\ref{E:coulcont}) become

\begin{equation}\label{E:outcont1}
v_{NC}^{(1)}(y)= \sin\Psi(y)\cos\Omega  +
c_N q \cos\Psi(y)\sin\Omega  ,
\end{equation}
while the function $v_{NC}^{(2)}$ can be obtained from eq. (\ref{E:outcont1})
by replacing $\sin\Psi\leftrightarrow\cos\Psi $ and $c_N$ by $-c_N$. In eq.
 (\ref{E:outcont1})

$$
\Psi(y)= q\ln 2ky +\frac{\pi}{2} +2qC + \frac{1}{2}q\left[\psi\left(1+{\rm i}\frac{qE}{pk}\right)
+ \psi\left(1 - {\rm i}\frac{qE}{pk}\right)   \right],~ c_N =\frac{\pi}{2}\left(1 + \coth \frac{\pi q}{\varphi_{N}}\right).
$$

In the region $ky \gg 1$ the vector $\vec{V}_{NC}$ (\ref{E:coulcont}), (\ref{E:outcont})
is determined by the components

\begin{equation}\label{E:outcont2}
v_{NC}^{(1)}(y)=2\frac{\tan\varphi_N}{q}
\exp \left(\frac{\pi}{2}\frac{qE}{pk}\right)\cos\frac{\varphi_N}{2}\sin\alpha(y)
\end{equation}
and $v_{NC}^{(2)}$, calculated from eq. (\ref{E:outcont2}) by replacing
$\cos\frac{\varphi_{N}}{2}$ by $\sin\frac{\varphi_{N}}{2}$ and $\sin\alpha$ by
$\cos\alpha$ with

$$
\alpha(y)= ky + \frac{qE}{pk}\ln 2ky  + \Omega.
$$

\emph{"Electric" region}
\\
\\
At the same time the "electric" two-vector

\begin{equation}\label{E:electr}
\vec{V}_{N\mbox{el}}(y)=\cos\vartheta\vec{V}_{N\mbox{el}}^{(1)}(y)
 + \sin\vartheta\vec{V}_{N\mbox{el}}^{(2)}(y)   ,
\end{equation}
where $\vartheta$ is an arbitrary phase,
is written in terms of the two-vectors $\vec{V}_{N\mbox{el}}^{(1,2)}$
calculated analogously to the vectors $\vec{V}_{N\mbox{el}}^{(+,-)}$
(\ref{E:elcomp}) incorporated into eq. (\ref{E:genel}). As a result the components
$v_{N\mbox{el}}^{(1,2)}$ of the vector $\vec{V}_{N\mbox{el}}$ (\ref{E:electr})
in the region $y \ll \tilde{y}_2$ become

\begin{equation}\label{E:electr1}
v_{N\mbox{el}}^{(1)}= \sin \left[\tilde{S}(y) +\frac{\pi}{4} + \vartheta \right];\quad
v_{N\mbox{el}}^{(2)}= \tan\frac{\varphi_{N}}{2}
\cos \left[\tilde{S}(y) +\frac{\pi}{4} + \vartheta \right],
\end{equation}
where

$$
\tilde{S}(y)= \frac{(ka_0)^3}{3f} +ky +\frac{qE}{pk}\ln\frac{y}{4\tilde{y}_2}.
$$

On matching the wave-vectors $\vec{V}_{N\mbox{it}}$ (\ref{E:iter}) and $\vec{V}_{N\mbox{C}}$
(\ref{E:coulcont}) in
the intermediate region $ d\ll y \ll k^{-1} $ we obtain for the phase $\Omega$

\begin{equation}\label{E:cotcont}
\cot \Omega = - \frac{c_N}{T(k)},
\end{equation}
where

\begin{equation}\label{E:T}
T(k) = \ln \frac{kD}{2} +2C-1 + \frac{1}{2}\left[\psi\left(1+{\rm i}\frac{qE}{pk}\right)
    + \psi\left(1  - {\rm i}\frac{qE}{pk}\right)   \right].
\end{equation}

A comparison in the other intermediate region $k^{-1} \ll y \ll \tilde{y}_2$ the wave vectors
$\vec{V}_{N\mbox{C}}$ (\ref{E:outcont2}) and $\vec{V}_{N\mbox{el}}$ (\ref{E:electr1})
yield

\begin{equation}\label{E:link}
\vartheta = \Omega - \frac{(ka_0)^3}{3f}
 +\frac{qE}{pk}\ln 4\frac{(ka_0)^3}{f} - \frac{\pi}{4}.
\end{equation}

Equation (\ref{E:cotcont}) allows to calculate the phase $\Omega$ as a function
of the energy $E$. As expected setting $k= {\rm i}\frac{\varepsilon_N q}{p\eta}$
and matching the functions (\ref{E:coulcont}), (\ref{E:outcont}) taken
at $|t|\ll 1$ with the iteration functions (\ref{E:iter}) and then
with the "electric" functions (\ref{E:electr1}) at $|t|\gg 1$,
we obtain the equation (\ref{E:cot})
for the $\cot\Omega_0$. Employing the equation $\cot\Omega_0 = {\rm i}$,
determining the poles of the $S$-matrix $\left(S = \exp (2{\rm i}\Omega)  \right)$
\cite{landau, berlif}, we arrive at eq. (\ref{E:compleq}) for the quasi-discrete
energy levels. Note, that the wave-vector $\vec{V}_{N\mbox{el}}$ (\ref{E:electr})
has at $y \gg \tilde{y}_2$ the asymptotic form of the standing wave with
the components

 \begin{eqnarray}
 v_{N\mbox{el}}^{(1)}(y)&=&
 \sin \left[\frac{2^{2/3}}{3} \left(\frac{F}{F_0}\right)^{1/2} \left(\frac{y}{a_0}\right)^{3/2}
 +\frac{\pi}{4} +\vartheta  \right];
 \nonumber\\
 v_{N\mbox{el}}^{(2)}(y)&=& \tan\frac{\varphi}{2}
 \cos \left[\frac{2^{2/3}}{3} \left(\frac{F}{F_0}\right)^{1/2} \left(\frac{y}{a_0}\right)^{3/2}
 +\frac{\pi}{4} +\vartheta  \right];
 \end{eqnarray}

 The main result (\ref{E:cotcont}) of this subsection is valid under the conditions $q\ll 1,~\frac{pk}{E}\ll 1$
 and $f(ka_0)^{-3} \ll 1$.

 \section{Double-subband approximation}\label{S:double}

In this section we consider the coupling between the continuous states branching
from the ground size-quantized energy level $\varepsilon_0$ and discrete states
adjacent to the energy level $\varepsilon_1$, having the common energies
$E = \sqrt{\varepsilon_0^2 + p^2 k^2}= \varepsilon_1\sqrt{1 - \frac{q^2}{\eta^2}}$.
The corresponding four-fold set can be derived from the set (\ref{E:set})
limited by $N,N' = 0,1$.
\\
\\
\emph{Continuous states } $N=0$
\\
\\
In the inner region $d \ll y \ll k^{-1}$ the above described iteration procedure leads to the
components $v_{0\mbox{it}}^{(1,2)}$ of the vector $\vec{V}_{0\mbox{it}}$ \cite{monschm12}

\begin{eqnarray}\label{E:set0}
\left.
\begin{array}{c}
v_{0 it}^{(1)}(y)=R_0\sin (Q(y) + \zeta_0) + R_1 q \gamma_{01}\cos\zeta_1~;\\
v_{0 it}^{(2)}(y)=R_0\cos (Q(y) + \zeta_0) - R_1 q \gamma_{01}\sin\zeta_1~;
\end{array}
\right \}
\end{eqnarray}
In this set $Q(y)$ is given by eq. (\ref{E:iter}), $R_{0,1}$ and $\zeta_{0,1}$
are the corresponding arbitrary constants and phases, respectively. The parameter

\begin{eqnarray}\label{E:gamma01}
\pi\gamma_{01}&=&\cos\alpha_0 \left[  {\rm {Ci}}\left( \frac{\pi}{2} +\alpha_0\right)-
{\rm {Ci}}\left( \frac{\pi}{2} -\alpha_0\right) \right]
\nonumber\\
&+&\sin\alpha_0 \left[  {\rm {Si}}\left( \frac{\pi}{2} +\alpha_0\right)+
{\rm {Si}}\left( \frac{\pi}{2} -\alpha_0\right) \right],~\alpha_0=\frac{\pi x_0}{d}.
\end{eqnarray}
consisting of the integral sine ${\rm {Si}}~ \mbox{and cosine}~ {\rm {Ci}}$ \cite{abram},
describes the coupling induced by the potentials $V_{01}=V_{10}$ (\ref{E:pot}). In this
region the components $v_{0\mbox{C}}^{(1,2)}$ of the Coulomb vector $\vec{V}_{0\mbox{C}}$
can be calculated from eq. (\ref{E:outcont1}) for $N=0$.

In the "electric" region $ k^{-1} \ll y \ll \tilde{y}_2 $ the components $v_{0\mbox{C}}^{(1,2)}$
of the Coulomb vector $\vec{V}_{0\mbox{C}}$ coincide with those presented in eq.
(\ref{E:outcont2}), while the wave functions $v_{0\mbox{el}}^{(1,2)}$ relevant
to the "electric" vector $\vec{V}_{0\mbox{el}}$ are given by eq. (\ref{E:electr1}).

Matching in the inner region the wave vectors $\vec{V}_{0\mbox{it}}$
(\ref{E:set0}) with $\zeta_0 = \zeta_1 = \frac{\pi}{2}$
 and $\vec{V}_{0\mbox{C}}$ (\ref{E:outcont1}) we arrive at

\begin{equation}\label{E:setcoupl}
R_0 [T(k) \cot \Omega + c_0] - R_1\gamma_{01}\cot \Omega = 0,
\end{equation}
where $T(k)$ is defined in eq. (\ref{E:T}). On equating in the "electric"
region the Coulomb vector $\vec{V}_{0\mbox{C}}$ (\ref{E:outcont2})
and "electric" vector $\vec{V}_{0\mbox{el}}$ (\ref{E:electr1})
the relationship (\ref{E:link}) between the phases $\Omega$ and $\vartheta$
of the Coulomb and "electric" wave-vectors, respectively, is obtained.
\\
\\
\emph{Discrete states} $N=1$
\\
\\
In the inner region $d \ll  y \ll a_0$ the components $v_{1\mbox{it}}^{(1,2)}$
of the wave-vector $\vec{V}_{1\mbox{it}}$ are obtained from the wave functions
$v_{0\mbox{it}}^{(1,2)}$ (\ref{E:set0}), respectively by replacing
$R_0\leftrightarrow R_1$ and $\zeta_0 \leftrightarrow\zeta_1$. The Coulomb wave-vector
$\vec{V}_{1\mbox{C}}$ is defined by the components $v_{1\mbox{C}(-)}^{(1,2)}$
(\ref{E:outer1}) and $v_{1\mbox{C}(+)}^{(1,2)}$ (\ref{E:outer2}) of the wave-vectors
$\vec{V}_{1\mbox{C}}^{(+,-)}$ in eq. (\ref{E:coulvec}). In the "electric"
region $a_0 \ll y \ll \tilde{y}_2$ the corresponding wave functions
$v_{1\mbox{C}(-)}^{(1,2)}$ and $v_{1\mbox{C}(+)}^{(1,2)}$ have the form
(\ref{E:outer3}) and (\ref{E:outer4}), respectively. The "electric"
wave vector

$$
\vec{V}_{1\mbox{el}} = C_1\left[\sin\vartheta\vec{V}_{1\mbox{el}}^{(-)} +
  \cos\vartheta \vec{V}_{1\mbox{el}}^{(+)}    \right]
$$
formed by the vectors $\vec{V}_{1\mbox{el}}^{(+,-)}$, having the components
(\ref{E:elcomp}) for $N=1$, gives for the $v_{1\mbox{el}}^{(1,2)}$

\begin{eqnarray}\label{E:elreal}
\left.
\begin{array}{c}
v_{1 \mbox{el}}^{(1)}(y)=C_1 \pi^{-\frac{1}{2}}x_2^{-\frac{1}{4}}
 \left(\sin\vartheta\exp \left[S(y)   \right]+
\frac{1}{2}\cos\vartheta\exp \left[- S(y)   \right]    \right)~;\\
v_{1 \mbox{el}}^{(2)}(y)=C_1 \pi^{-\frac{1}{2}}x_2^{-\frac{1}{4}}
\frac{1}{2} \frac{q}{\eta}
\left(-\sin\vartheta\exp \left[S(y)   \right]+
\frac{1}{2}\cos\vartheta\exp \left[- S(y)   \right]    \right)~;
\end{array}
\right \}
\end{eqnarray}
with eq. (\ref{E:momas}) for $S(y)$.

On equating in the inner region the wave vectors $\vec{V}_{1\mbox{it}}$
 and $\vec{V}_{1\mbox{C}}$ calculated from eqs. (\ref{E:set0}) and
(\ref{E:coulvec}), (\ref{E:outer1}), (\ref{E:outer2}), respectively, we obtain

\begin{eqnarray}\label{E:eqel1}
R_{1} + R_{-}\frac{1}{\eta^2\Gamma (-\eta)}=0
\nonumber\\
R_{1}Y(\eta) - R_{+}\frac{1}{\eta} - R_0 \gamma_{01}=0
\end{eqnarray}
where $Y(\eta)$ is given by eq. (\ref{E:Y1}).

A comparison in the "electric" region for the Coulombic   $\vec{V}_{1\mbox{C}}$
(\ref{E:coulvec}), (\ref{E:outer3}), (\ref{E:outer4}) and "electric"
$\vec{V}_{1\mbox{el}}$ (\ref{E:elreal}) wave vectors leads to the set

\begin{eqnarray}\label{E:eqel2}
-R_{-} + &R_{+}&\frac{(-1)^{\eta}}{\Gamma (\eta)} +
 C_1\pi^{-\frac{1}{2}}x_2^{-\frac{1}{4}} \eta \sin\vartheta \Phi_{\eta}^{-1}=0
\nonumber\\
&R_{+}&\frac{1}{\Gamma (-\eta)} +
 C_1\pi^{-\frac{1}{2}}x_2^{-\frac{1}{4}} \eta \sin\vartheta \frac{\cot\vartheta}{2}\Phi_{\eta}=0
\end{eqnarray}
with eq. (\ref{E:Phi}) for $\Phi_{\eta}$. The total set of eqs.
(\ref{E:setcoupl}), (\ref{E:eqel1}) and (\ref{E:eqel2}) for the coefficients
$R_{+,-},~R_{0,1}$ and $C_1$ being solved by the determinantal
method gives

\begin{equation}\label{E:cotcoupl}
\left[\frac{(-1)^{\eta}\sin \pi\eta}{\pi}+
\frac{2\eta}{\Gamma^2 (1-\eta)}\frac{\Phi_{\eta}^{-2}}{\cot\vartheta}\right]
 \left[Y(\eta)-\frac{\gamma_{01}^2}{T(k)+ \frac{c_0}{\cot\Omega}}\right] -1 = 0,
\end{equation}
where $Y[\eta(E)]$ (\ref{E:Y1}), $T[k(E)]$ (\ref{E:T}), $c_0[k(E)]$ (\ref{E:outcont1})
are introduced above. The phases $\Omega$ and $\vartheta$ are linked by
eq. (\ref{E:link}). By solving eq. (\ref{E:cotcoupl}) the phases $\Omega$
and $\vartheta$ as a function of the energy $E$ can be found in principle.

As expected the general eq. (\ref{E:cotcoupl}) describes the limiting
cases studied above for negligibly small coupling
 $(\gamma_{01}\rightarrow 0)$ or electric field $(\Phi_{\eta}\rightarrow 0)$.
 Equation $\cot \vartheta = {\rm i}$ with $\cot \vartheta$ calculated from
 eq. (\ref{E:cotcoupl}) at $\gamma_{01}=0$ coincides with
 eq. (\ref{E:compleq}) derived in the approximation of the isolated subbands.
 Equation $\cot \Omega = {\rm i}$ with $\cot \Omega$ taken from eq. (\ref{E:cotcoupl})
 at $\Phi_{\eta}=0$ transforms into that describing the Fano resonances in the
 double-subband approximation \cite{monschm12}.

 The complete energy width $\Gamma_{1n}^{(\mbox{c})}$ caused by both mechanisms
 of ionization can be derived by setting $\cot \vartheta = {\rm i}$ in eqs. (\ref{E:link})
 and (\ref{E:cotcoupl}) and then expanding $Y(\eta)$ (\ref{E:Y1}) in the vicinity of
 the quantum numbers $\eta_{1n}$ of the strictly discrete states for which
 $Y(\eta_{1n})=0$. The complex quantum numbers $\eta_{1n}(E)$ calculated from
 eq. (\ref{E:cotcoupl}) determine the quasi-discrete energy states

 \begin{equation}\label{E:encompl}
 E_{1n} = \varepsilon_1\left(1 - \frac{q^2}{2\eta^2}  \right) -
  {\rm i}\frac{\Gamma_{1n}^{(\mbox{c})}}{2}~;
 \end{equation}
including the complete energy width

 \begin{equation}\label{E:sum}
 \Gamma_{1n}^{(\mbox{c})} = \Gamma_{1n}^{(\mbox{el})} + \Gamma_{1n}^{(\mbox{F})}~;
 \end{equation}
 where the width $\Gamma_{1n}^{(\mbox{el})}$ induced by the electric field is given by
 eq. (\ref{E:width}). For the width $\Gamma_{1n}^{(\mbox{F})}$ of the Fano resonances
 we obtain

 \begin{eqnarray}\label{E:Fano}
\Gamma_{1n}^{(\mbox{F})}=
\left\{
\begin{array}{cl}
8(\arctan 2)^{-1}\delta_{10}^{-1}\varepsilon_1 q^3 \gamma_{01}^2~;\quad n=0~;
\\
\\
4(\arctan 2)^{-1}\delta_{1n}^{2}\varepsilon_1 q^3 \gamma_{01}^2~;\quad n=1,2,\ldots~;
\end{array}
\right.
\end{eqnarray}
where the quantum defects $\delta_{1n} = \eta - n~;$ \quad n=0,1,2,\ldots~
can be calculated from eq. $Y(\eta)=0$ using eq. (\ref{E:Y1}) at $N=1$ for $Y(\eta)$.

This point is suitable to demonstrate one of the possible applications of the obtained results.
Since the Breit-Wigner resonant scattering on the quasi-discrete state caused by the inter-subband coupling
has been considered in Ref. \cite{monschm12} below we focus on the effect of the resonant
capturing of the electron induced by the electric
field. The electron density within the Coulomb well is determined by the coefficient $R_{-}$
in the wave vector (\ref{E:coulvec}), growing towards the impurity centre. The electron density
$\sim |R_{-}|^2$ related to the ground size-quantized energy level
$\varepsilon_0$ can be obtained from eqs.
(\ref{E:eqel1}), (\ref{E:eqel2}), (\ref{E:cotcoupl})
at $\gamma_{01}=0$ and $N=0$. Using the function $Y(E)$ derived from eq. (\ref{E:compleq})
and then expanded in a series
in the vicinity of the resonant energy level $W_{0n}~(Y(W_{0n})=0)$ and the coefficient
$C_0$ providing the unit flux density of the waves in eq. (\ref{E:stand}), the electron density reads

\begin{equation}\label{E:dens}
|R_{-}|^2=\frac{1}{\eta^2-\frac{1}{2}\delta_{0n}^2}
 q^2\frac{\varepsilon_0}{\Gamma_{0n}^{(\mbox{el})}}w(\Delta E)~,
\end{equation}
where

$$
w(\Delta E)=\frac{\left(\frac{1}{2}\Gamma_{0n}^{(\mbox{el})}\right)^2}
{\left(\frac{1}{2}\Gamma_{0n}^{(\mbox{el})}\right)^2 + \Delta E ^2}~;\Delta E = E- W_{0n}~;\quad n=0,1,2,\ldots .
$$
In eq. (\ref{E:dens}) $w(\Delta E)$ is the probability of the resonant capturing
of the ingoing electron within the impurity region for a lifetime $\tau = \hbar /\Gamma_{0n}^{(\mbox{el})}$
before being transformed into an outgoing wave. Eq. (\ref{E:dens}) is completely in line
with the results obtained for the $\delta-$function \cite{Mello} and 3D rectangular \cite{Baz}
potential barriers.

\section{Discussion}\label{S:Disc}

\emph{Single subband approximation}
\\
\\
We define the binding energy of the electron $E_{Nn}^{(b)}$ of the impurity electron
in the $n$ quasi-discrete
state associated with the $N$ subband as the difference between
the size-quantized energy $\varepsilon_N$ (\ref{E:ener}) of the free electron
and the real part of the
the energy of the impurity electron $E_{Nn}$ given by eqs. (\ref{E:elen}), and
(\ref{E:encompl}),
yielding
$E_{Nn}^{(b)}=\varepsilon_N q^2/2\eta^2,~\eta = n + \delta_{Nn},~n=0,1,2,\ldots $.
The dependencies of the binding energy on
the ribbon width $d$ and the displacement of the impurity centre $x_0$ from the
mid-point of the ribbon $x=0$ were discussed in detail in Ref. \cite{monschm12}.
Here we only mention that the binding energy decreases with increasing
ribbon width $d~(E_{Nn}^{(b)} \sim \varepsilon_N \sim d^{-1} )$ and with
shifting the impurity from the ribbon centre towards the boundaries. Note,
that we ignore the small effect of the electric fields on the binding energy.
In order to calculate the corrections $\Delta E_{N0}$ and $\Delta v_{N0}$
to the non-relativistic energy $E_{N0}$ (\ref{E:elen}) and the wave function
$v_{N0}(\tau)\sim \exp (-\frac{\tau}{2})$ (\ref{E:outer}), respectively, caused by the
electric field $f=F/F_0$, we trivially solved the equation

$$
v''(\tau) + \frac{\eta^3}{4}f\tau v(\tau) -
 \frac{\eta^2 a_0}{2pq}E_{N0}^{(b)}v(\tau) = 0,
$$
by setting $\Delta v\sim f$ and $\Delta E_{N0}^{(b)}\sim f^2$
to find

$$
\Delta E_{N0}^{(b)} =E_{N0}^{(b)}\frac{5}{4}\delta_{N0}^6 f^2.
$$
The obtained red shift of the energy level $E_{N0}$ coincides completely
with that calculated by Ratnikov and Silin \cite{RatSil} by the
Dalgarno-Lewis perturbation theory method \cite{DalgLew}. For the GNR
of width $d=2~\mbox{nm}$ placed on the sapphire substrate $(q=0.24)$
and exposed to the electric field $F\simeq F_0 \simeq 20~\mbox{kV/cm}$
the relative shift of the binding energy of the ground impurity state
$\Delta E_{N0}^{(b)}/E_{N0}^{(b)} \simeq 5.2\cdot 10 ^{-3} $ is
negligibly small.

\begin{figure}[htbp]
\begin{center}
\includegraphics[width=8.6cm]{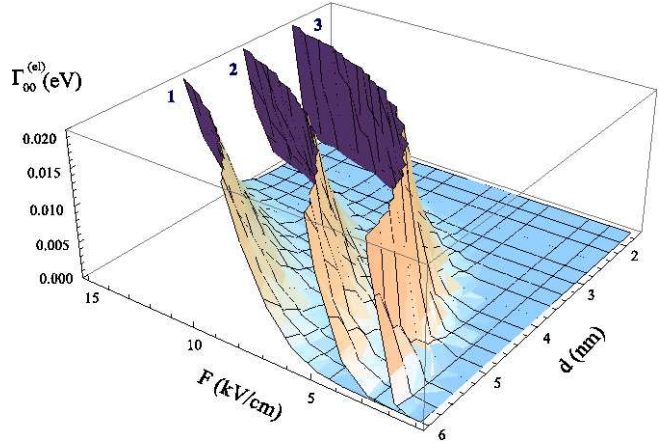}
\end{center}
\caption{\label{qd2} (color online). The width $\Gamma_{00}^{(\mbox{el})}$ (\ref{E:width}) of the ground impurity
state $n=N=0$ in the GNR placed on the $\mbox{SiO}_{2}$ substrate $(q=0.37)$
as a function of the electric field $F$ and of the graphene width $d$ for the different
impurity positions $x_{0}=s\frac{d}{2} $ with $s= 0.0-(1), 0.5-(2), 0.7-(3)$.}
\end{figure}

The main effect of the electric field is the ionization of the impurity
states which are accompanied  by the emergence of the energy widths $\Gamma_{Nn}^{(\mbox{el})}$.
It follows from eq. (\ref{E:width}) that
with increasing ribbon width $d$ and strength of the electric fields
$F$ the width of the quasi-discrete impurity states increases.
However, the greater the shift of the impurity centre $x_0$ from the mid-point
$x=0$ is the wider the impurity state becomes. This
means that in contrast to quasi-1D semiconductor structures
(QWR, bulk material subject to a magnetic field) in which the ionization
of the impurity centre is reached only by the increasing electric field,
in the GNR the mechanism of the dimensional ionization can be realized. The electric
field could be kept constant, while the widening of the ribbon and the displacement
of the impurity would lead to the ionization. Note, that the dimensional ionization is more
efficient as compared to the electric ionization, because the argument of the
exponent function in eq. (\ref{E:Phi}) changes $\sim d^{-2}$ with changing the ribbon
width $d$, and $\sim F^{-1}$ with changing the electric field $F$.
The width $\Gamma_{00}^{(\mbox{el})}$ of the ground impurity state adjacent
to the ground size-quantized level $N=0$ as a function of the ribbon width $d$
and electric field $F$ for the different impurity positions $x_0$ is depicted in Fig.2.
Iso-width lines $F\sim d^{-2}$ providing the width
$\Gamma_{00}^{(\mbox{el})}(F,d;x_0)=\mbox{const.}~(x_0= \mbox{const.})$
and given in Fig.3 evidently follow from eq. (\ref{E:width}).

\begin{figure}[htbp]
\begin{center}
\includegraphics[width=8.6cm]{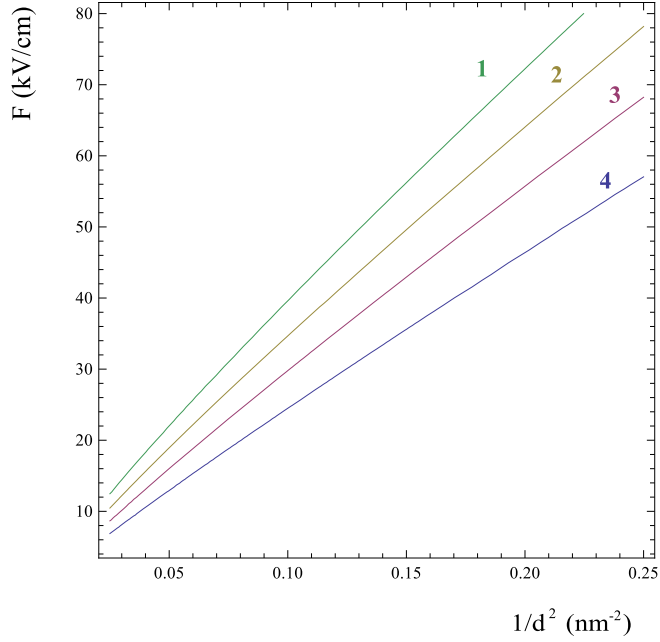}
\end{center}
\caption{\label{qd3} (color online). The iso-width curves $\Gamma_{00}^{(\mbox{el})} (d,F)=\mbox{const.}$ calculated from eq.
(\ref{E:width}) for the ground impurity state $(n=N=0)$
in the GNR placed on a $\mbox{SiO}_{2}$ substrate $(q=0.37)$. The impurity
is positioned at the mid-point $x_{0}=0 ~(s=0)$ of the GNR of width $d$; $F$ is the
strength of the electric field. The energy widths are taken to be
$\Gamma_{00}^{(\mbox{el})} = 0.10-(1),0.060-(2),0.030-(3),0.010-(4)~\mbox{eV}$.}
\end{figure}

 Fig.4 demonstrates
the iso-width surfaces when all parameters $F,d,x_0$ are changed.
In the GNRs the effects of both
 parameters $F$ and $d$ are governed by the exponential factor (\ref{E:Phi})
 in contrast to the semiconductor QWR in which the dependence on the radius
 of the QWR $R$ has the less pronounced logarithmic character \cite{monschm09}.
Equations (\ref{E:width}) and (\ref{E:Phi}) show that the impurity electron becomes
practically unbound if the electric field $F$ exceeds the critical value
$F^{(\eta)}\simeq F_0/\eta^3$. Thus, the ground state $(\eta = \delta_{N0})$
is less sensitive to the ionizing effect of the electric field and remains
stable up to the significantly greater electric fields $F^{(0)}$ than those
$F^{(\eta)}$, destroying the excited states $(\eta \simeq 1,2,\ldots)$.

\begin{figure}[htbp]
\begin{center}
\includegraphics[width=8.6cm]{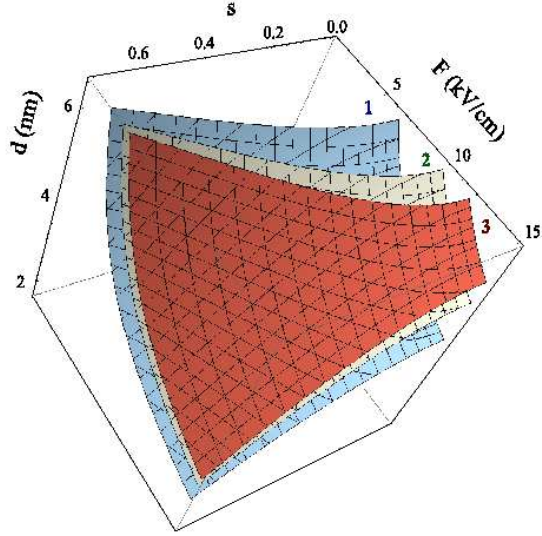}
\end{center}
\caption{\label{qd4} (color online).  The iso-width surfaces $\Gamma_{00}^{(\mbox{el})} (d,F,s)=\mbox{const.}$ calculated from eq.
(\ref{E:width}) for the ground impurity state $(n=N=0)$
in the GNR of width $d$ placed on the $\mbox{SiO}_{2}$ substrate $(q=0.37)$
and exposed to the electric field $F$. The impurity centre is positioned at the
coordinate $x_{0}=s\frac{d}{2} $. The energy widths are taken to be
$\Gamma_{00}^{(\mbox{el})} = 0.010-(1),0.060-(2),0.10-(3) ~\mbox{in eV}$.}
\end{figure}

\emph{Double subband approximation}
\\
\\
The double subband $N=0,1$ approximation describes the
combined effect of the electric field
ionization and inter-subband autoionization. Since the influence of the
electric field was discussed just above here we briefly remind the reader of the sequences
of the inter-subband interaction.
At $F=0$ the Rydberg series of the strictly discrete energy levels
adjacent to the excited size-quantized energies $\varepsilon_N ,~ (N\neq 0)$
transform into the quasi-discrete levels (Fano resonances)
of widths proportional to $\varepsilon_1 \sim d^{-1}$
and increasing both with decreasing the ribbon width $d$ and with the displacement
of the impurity centre from the mid-point of the ribbon $x=0$ \cite{monschm12}.

The combined effect of the both types of the ionization reflected in eqs.
(\ref{E:link}) and (\ref{E:cotcoupl}) leads to the summation of the widths
$\Gamma_{Nn}^{(\mbox{el})}$ and $\Gamma_{Nn}^{(\mbox{F})}$, associated
with the electric filed and Fano mechanisms, respectively (\ref{E:sum}).
The energy widths $\Gamma_{10}^{(\mbox{el})}$ and $\Gamma_{10}^{(\mbox{F})}$
as a function of the width $d$ for the different strengths $F$ of the electric field
and for the impurity positioned at $x_0 =d/4$ are presented in Fig.5.

\begin{figure}[htbp]
\begin{center}
\includegraphics[width=8.6cm]{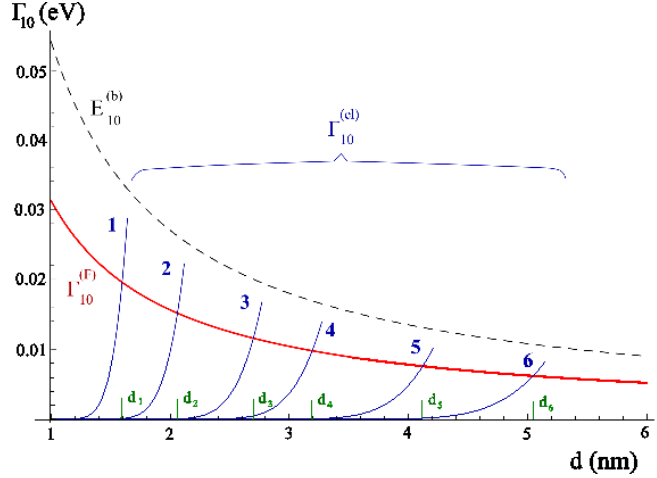}
\end{center}
\caption{\label{qd4a} (color online).  The dependence of the widths $\Gamma_{10}^{(\mbox{el})}$ (\ref{E:width})
and $\Gamma_{10}^{(\mbox{F})}$ (\ref{E:Fano}) of the ground state $n=0$
of the first excited size-quantized Rydberg series $N=1$, caused by the electric field
$F$ and the inter-subband coupling, respectively, and the binding energy
$E_{10}^{(b)}=\varepsilon_{1} - E_{10} $ on the width $d$ of the GNR placed
on the sapphire substrate $q=0.24$. The impurity centre is positioned at the coordinate
$x_{0}=0.5\frac{d}{2}$. The electric fields are taken to be
$F=2 - (1), 3 - (2),  5 - (3),  7 - (4),  12 - (5), 20 - (6)~\mbox{kV/cm}$.
The corresponding critical ribbon widths $d_F$ provides the balance between
the widths $\Gamma_{10}^{(\mbox{el})}$ and $\Gamma_{10}^{(\mbox{F})}$.}
\end{figure}

Clearly, $\Gamma_{Nn}^{(\mbox{el})}$ and $\Gamma_{Nn}^{(\mbox{F})}$
change with changing $d$ in the opposite way. As a result
in narrow GNRs the widening effect of the Fano coupling
exceeds that of the electric field, but with increasing the ribbon width
both effects come into balance and then the electric field ionization
dominates the autoionization. The greater is the electric field $F$
the less the critical width $d_F$ becomes, providing the equality between
the both widths. The parameters $d_F$ and $F$ obey the relationship
$d_F^2 F =\mbox{const.}$, following from the condition
$\Gamma_{Nn}^{(\mbox{el})}=\Gamma_{Nn}^{(\mbox{F})}$, in which the widths
$\Gamma_{Nn}^{(\mbox{el})}$ and $\Gamma_{Nn}^{(\mbox{F})}$ are given by eqs.
(\ref{E:width}) and (\ref{E:Fano}), respectively for $N=1, n=0$.

Since the Fano coupling does not contribute to the most interesting ground
impurity series $N=0$ and does not manifests itself
in not significantly narrow GNRs exposed to sufficiently strong electric
fields, we clarify below the mechanism of the resonant capturing of the electron
by the impurity centre in the presence of the electric fields. Eqs.
(\ref{E:eq1}), (\ref{E:eqel}) show that the ratio $R_{+}/R_{-}= Y/\Gamma (1-\eta)$
for the energy of the ingoing electron $E$ apart from the resonant value $W_{Nn}$
(arbitrary $\eta$ and $Y\sim O (\ln q)$ ) reads $|R_{+}/R_{-}| \gg 1$. The ingoing
wave then almost entirely reflects from the barrier. However, under the condition
$E\simeq W_{Nn}$ implying $\eta = n + \delta_{Nn}~, n=0,1,2,\ldots $, eqs.
(\ref{E:eq1}), (\ref{E:eqel}) and (\ref{E:compleq}) result in
$Y\sim O (\Phi_{Nn}^2 /\delta_{Nn})$, $|R_{+}/R_{-}| \ll 1$ for the
$|R_{-}|^2$ given by eq. (\ref{E:dens}). In case of the exact resonance
$(\Delta E = 0, w(0)=1)$ the electron density reaches a maximum
$$
|R_{-}|^2_{\mbox{max}} \sim q^2 \varepsilon_N /\Gamma_{Nn}^{\mbox{(el)}}\gg 1,
$$
while for the energy deviations considerably exceeding the resonant width
$\Delta E \gg \Gamma_{Nn}^{\mbox{(el)}}$ the electron density reduces
relatively $|R_{-}|^2_{\mbox{max}}$ by a factor
$w(\Delta E) = \left(\Gamma_{Nn}^{\mbox{(el)}} / 2 \Delta E \right )^2 \ll 1$.
Note, that the probability of the resonant capturing is very sensitive to
the accuracy of the resonant energy $W_{Nn}$. Neglecting the deviation of the potential
$V_{NN}(y)$ from the Coulomb form $V_{NN}(y) \sim -|y|^{-1}$ (\ref{E:pot}) at small
distances $|y| \ll d$ for which $V_{NN}(y) \sim \ln (|y|/ d) $ \cite{monschm12} and setting
$\delta_{Nn}=0,~ \eta = n \neq 0$, induce the energy shift
$\Delta E = \sim q^2 \varepsilon_N n^{-3}\delta_{Nn}$. This shift significantly exceeding the resonant
width $\Gamma_{Nn}^{\mbox{(el)}}$ results in
$\Gamma_{Nn}^{\mbox{(el)}} / \Delta E \sim \mbox{O}(\delta_{Nn}^{-1}\Phi_{Nn}^2)\ll 1$.
In conclusion of this paragraph note that the specific problems of the electron
scattering on the impurity centres in GNRs having the signs of the resonant and
potential scattering require special consideration.
\\
\\
\emph{Estimates of the expected experimental values}
\\
\\
In an effort to render our results close to an experimental setup, we present below
the estimates of the expected values for the GNRs of the family corresponding to
$\tilde{\sigma}=\frac{1}{3}$ placed on $\mbox{SiO}_{2}~(\epsilon =3.9, q=0.37)$ and
sapphire $(\epsilon =10, q=0.24)$ substrates \cite{han}. Since the $\mbox{SiO}_{2}$
material is not the best candidate to be described by the theory implying
$q\ll 1$ the general equation $E^2 (1+q^2/\eta^2)=\varepsilon_{N}^{2}$ for the energy $E$
has to be taken to calculate the binding energy $E_{N\eta}^{(b)}$, width $\Gamma_{N\eta}$,
electric field $F^{(\eta)}$ and other parameters. Being derived from this equation
and from eq. $Y(\eta)=0$ (\ref{E:Y}) the binding energy of the ground Rydberg state
$n=0$ of the ground size-quantized series $N=0$ for the impurity positioned at the
mid-point $(x_{0}=0)$ of the GNR of width $d=2~\mbox{nm}$ reads $E_{00}^{(b)}=68~\mbox{meV}$.
For the critical electric field $F^{(\eta)}$ providing the complete depletion
of the $\eta$ impurity level and estimated from the condition
$eF^{(\eta)} a_{0}\eta \simeq E_{N\eta}^{(b)}$ we obtain $F^{(0)}=450~\mbox{kV/cm}$.
The less bound first excited impurity level $(E_{01}^{(b)} \simeq 54~\mbox{meV})$ can
be ionized by the electric field $F^{(1)}=310~\mbox{kV/cm}$. The above mentioned condition related to the
fields $F^{(\eta)}$
is suitable
to introduce the parameter of stability $Q_{Nn}(q,s)$ of the $n$ impurity state
associated with the $N$ subband relatively to the ionization effect of the electric
 field $F$

 \begin{eqnarray}
 Q_{Nn}(q,s)&=& \frac{\pi^{2} p}{e} |N-\tilde{\sigma}|^{2}G_{Nn}(z);~
 \frac{\pi^{2} p}{e}=6.53~\mbox{Vnm};
 \\
 G_{Nn}(z)&=&2z\frac{[\sqrt{z^{2}+1} -1]}{z^{2}+1};
 \nonumber
 \end{eqnarray}
where $z \equiv z_{Nn}(s)< 1$ is the root of the equation $Y(z_{Nn})$ (see eq.(\ref{E:Y})).
Under the condition $Fd^{2} > Q_{Nn}$ the $Nn$ state is practically ionized, while in the opposite
case $Fd^{2} < Q_{Nn} $ the given state can be treated as relatively stable.
 For the possibly employed substrates, namely $\mbox{SiO}_{2}$ $(q=0.37)$,
sapphire $(q=0.24)$ and $\mbox{HfO}_{2}$ $(q=0.13)$ the corresponding parameters calculated for the
impurity positioned at the ribbon mid-point $(s=0)$ read
$Q_{00}=0.18,~ 0.12, ~\mbox{and}~ 0.063~\mbox{Vnm}$, respectively.

In order to estimate the combined effect of the electric field and Fano-ionization
on the impurity states adjacent to the $N=1$ subband we are forced to avoid the
ribbon placed on the $\mbox{SiO}_{2}$ substrate and address a sapphire
substrate. The point is that the condition of the adiabatic approximation
$a_0 \ll d$ with eq. (\ref{E:par}) for $a_0$ being written strictly looks like
$\pi |N-\tilde{\sigma}|q \ll 1$, which, as pointed above, transforms into eq.
(\ref{E:weakimp}) for the low excited $N$ subbands. For the chosen
$\tilde{\sigma} =1/3$ the ground subband $N=0$ provides for the
$\mbox{SiO}_{2}$ substrate $\pi |N-\tilde{\sigma}|q \simeq q \simeq 0.37 $,
while for the subband $N=1$ this parameter is already $2q = 0.74$ that makes
the adiabatic approximation for this subband
for the $\mbox{SiO}_{2}$ substrate to be inappropriate. It follows from Fig. 5
that the resonant Fano width consists to a considerable part of the binding energy of the ground state
$(\Gamma_{10}^{(\mbox{F})}\simeq \frac{2}{3}E_{10}^{(b)} )$.
The possible reasons for this are first the parameter $q=0.24$
being close to the threshold of the adiabatic approximation $2q \ll 1$ and second
the ground state $n=0, \delta_{10} < 1$ is more sensitive to the Fano-coupling
$\Gamma_{10}^{(\mbox{F})} /E_{10}^{(b)} \sim q\delta_{10} $ than the excited states
$n= 1,2,\ldots, \Gamma_{n1}^{(\mbox{F})} /E_{1n}^{(b)} \sim q\delta_{1n}^{2}/n,~ \delta_{1n}< 1$.
The excited states $n\neq 0$ are expected to be significantly narrower than
the ground state $n = 0$.

In the presence of relatively weak electric field
$F< 7~\mbox{kV/cm}$ the lifetime $\tau_{10}=\hbar / \Gamma_{10}^{(\mbox{F})}$  of the state
$n=0, N=1$ in the ribbon of width $d=2~\mbox{nm}$ is determined only by the
Fano width $\Gamma_{10}^{(\mbox{F})} \simeq 0.015~\mbox{eV} $
resulting in $\tau_{10} = 4.4\cdot 10^{-14}~\mbox{s}$. However the lifetime of the
first excited state $n=N=1$ in the same ribbon $\tau_{11}=\hbar / \Gamma_{11}^{(F)}$
is of the order of $\tau_{11} = 1.0\cdot 10^{-12}~\mbox{s}$.
Thus even in the absence of electric field the resonant Fano widths of the impurity states
should be taken into account in the study of the electronic and transport processes in GNRs.
Recently Gong \emph{et.al.} \cite{gong} reported that the analogous line-defect-induced
Fano resonant states in the conduction band of the armchair GNR impede the electron
transport in this region.
With increasing electric field $F$ and decreasing the ribbon width $d$
the contribution of the electric field to the complete width becomes more pronounced.
The critical width $d_F$ at which the electric field $F=2,3,5,7,12,20~\mbox{kV/cm}$
~and the Fano coupling contribute equally to the energy width $\Gamma_{10}^{(\mbox{c})}$
are $d_{F} = 5.0,4.1,3.2,2.7,2.0,1.6~\mbox{nm}$. The dependence $d_F \sim F^{-1/2}$
is valid to a high accuracy. It should be noted that at the critical ribbon widths $d_F$
the ground impurity state $n=0$ in the ribbon located on the chosen specific substrate
$q=0.24$ seems to be completely ionized. At the same time the substrates with the
greater dielectric constant $\epsilon$ (the less $q$) provide significantly stable
impurity states especially those having the quantum numbers $n>1$.

A comparison of our results with those obtained numerically based on density
functional theory \cite{Zhu} and on the tight binding approximation \cite{Zheng},
\cite{Jia}, \cite{Mohamm} demonstrates that the Dirac equation approach employed
in this paper quite adequately describes the electronic structure and the
impurity and exciton states in the GNRs. The exciton characteristics can be
obtained from the corresponding impurity ones by replacing $p$ by $2p$ and $q$ by $\frac{1}{2}q$.
The dependencies $\sim d^{-1}$ of the effective electron mass
$M_{N}=| N-\tilde{\sigma} |\pi \hbar^{2}(pd)^{-1}$ \cite{monschm12},
the energy gap $E_{g}=2\varepsilon_{0}$ (\ref{E:ener}) and the binding energy
$E^{(b)}_{\mbox{exc}}= 1/2E^{(b)}_{\mbox{imp}}$ (\ref{E:elen}), (\ref{E:encompl})
on the ribbon width $d$ are qualitatively in line with those presented in all
above mentioned Refs. Moreover, the energy gaps $E_{g}$
reveal a quantitative good agreement. Thus, the energy gaps $E_{g}\simeq 0.68~\mbox{eV}$
and $E_{g}\simeq 0.89~\mbox{eV}$ calculated from (\ref{E:ener}) for $d=2~\mbox{nm}$
and $d=1.45~\mbox{nm}$, respectively are close to the values
$E_{g}\simeq 0.65~\mbox{eV}$ \cite{Zheng}
and $E_{g}\simeq 0.86~\mbox{eV}$ \cite{Mohamm} presented for the corresponding widths.
A greater discrepancy is found for the masses $M_{0}\simeq 0.072 $ of the electron in the ribbon
of width $d\simeq 1.5~\mbox{nm}$ scaled to the mass of the free electron $m_{0}$
and $M = 0.050$ \cite{Mohamm}. Though, the dependence $\sim d^{-1}$ of the binding energy
on the ribbon width $d$ correlates completely with that obtained numerically
\cite{Zhu}, \cite{Jia}, \cite{Mohamm}
the different environments prevent us from a detailed
quantitative comparison. This is because our data are calculated for the effective dielectric constant
$\epsilon_{\mbox{eff}}$ (\ref{E:coulomb}) resulting in $q\ll 1$, while others
for the
GNRs or suspended
$(q= 2.2)$ \cite{Zhu}, \cite{Jia} or placed on the $\mbox{SiO}_{2}$ substrate,
with unspecified dielectric constant $\epsilon_{\mbox{eff}}(r)$
(see eqs. (10) and (11) in paper \cite{Mohamm}) inducing $q\simeq O (1)$.
We therefore conclude that the presented analytical results well correlate with those
obtained by the numerical approaches in the literature.
Along with the estimates of the expected experimental values
this could be extended to further studies of the wide range of the GNR structures
and their applications in nanoelectronics.

\section{Conclusion}\label{S:Concl}

In summary, we have developed an analytical approach to the problem of
the impurity electron in a narrow armchair GNR exposed to the external electric field
directed parallel to the graphene axis. The effect of the strong confinement
is taken to be much greater than the influence of the impurity Coulomb
electric field, which in turn considerably exceeds the external field. In the
approximation of the isolated size-quantized subbands we have
calculated the complex energy levels of the quasi-discrete impurity Rydberg states
and phases of the wave functions of the continuous spectrum.
The complex energies determine the binding energies and widths of the quasi-stationary
states, while the phases (and S-matrix) allow to study the various scattering problems.
The explicit form of the obtained results makes it possible to trace the dependence of the
listed above values on all the parameters of the structure, namely, on the ribbon width,
position of the impurity centre, and the electric field. In particular
it was found that the GNR is the structure in which the mechanism
of the dimensional ionization occurs: the impurity centre can be ionized by
increasing the ribbon width. In the approximation of the ground and first excited size-quantized
subbands the complete widths of the first excited Rydberg series caused by the combined
effect of the electric fields and the Fano resonant inter-subband coupling
have been calculated. Estimates of the expected experimental values
for realistic GNGs  show that there are two aspects of the effect of the
electric field. Weak field provides the resonant capturing of the electrons
by the impurity centres for a significantly long lifetime, and remain quasi-discrete
impurity states available for the experimental
in particular optical study. Relatively strong field releases
the bound electrons to activate the transport properties of the GNRs.

\section{Acknowledgments}\label{S:Ackn}
The authors are grateful to D.B. Turchinovich for useful discussions
and A. Zampetaki for significant technical assistance.

\end{document}